# FIRST EXPERIENCES INTEGRATING PC DISTRIBUTED I/O INTO ARGONNE'S ATLAS CONTROL SYSTEM

F. H. Munson, D. E. R. Quock, S. L. Dean[*], and K. J. Eder[*],
ANL, Argonne, IL 60439, USA

Abstract

The roots of ATLAS (Argonne Tandem-Linac Accelerator System) date back to the early 1960's. Located at the Argonne National Laboratory, the accelerator has been designated a National User Facility, which focuses primarily on heavy-ion nuclear physics. Like the accelerator it services, the control system has been in a constant state of evolution. The present real-time portion of the control system is based on the commercial product Vsystem [1]. While Vsystem has always been capable of distributed I/O processing, the latest offering of this product provides for the use of relatively inexpensive PC hardware and software. This paper reviews the status of the ATLAS control system, and describes first experiences with PC distributed I/O.

## 1 ABOUT ATLAS

ATLAS is an ion accelerator that combines older accelerator technology with newer technology. The configuration of the accelerator system consists of three ion sources, two injectors, and two LINAC (Linear Accelerator) sections. One of the injectors is an electrostatic Tandem accelerator, which utilizes some of the earliest accelerator techniques. The second injector is a positive-ion LINAC, which uses some of the latest superconducting accelerator technologies. Ions from each injector are transported to the entrance of the first booster LINAC. Ions at the output of the first booster LINAC can then be transported to the entrance of the last booster LINAC, and then directed to one of several target areas. The LINAC portion of the accelerator consists of sixty superconducting accelerating resonator structures.

## 2 THE ATLAS CONTROL SYSTEM

### 2.1 System Hardware

An upgrade to the control system was completed in 1998 [2]. Like the accelerator, the control system employs both new and old technologies. The primary interface to various accelerator components is a CAMAC (Computer Automated Measurement And Control) subsystem. While CAMAC is still used extensively in some control and data acquisition systems, and while manufacturers continue to produce CAMAC equipment, CAMAC is considered by many to be an outdated technology. The CAMAC subsystem is configured as a single CAMAC Serial Highway, which currently links 18 crates, and operates at a clock speed of 2.5 MHz.

A second subsystem that is part of the control system is an Ethernet LAN (Local Area Network). This LAN is a hybrid system consisting of 10 MB/s and 100 MB/s segments. The Ethernet subsystem is used to link all of the control system's computers.

At the core of the control system is a Compaq AlphaServer [3]. This computer provides the only link to the CAMAC subsystem. This link is established by a PCI (Peripheral Component Interconnect) to CAMAC Serial Highway interface. The interface and a software driver were acquired from Kinetic Systems Inc. [4]. In addition to the online AlphaServer, the control system consists of a backup AlphaServer, five AlphaStations, and fourteen PCs.

### 2.2 System Software

The operating system software currently used by the ATLAS control system is Compaq's "OpenVMS" and Microsoft's "Windows NT/2000" [5].

Vista Control Systems' software package "Vsystem" is used to provide the real-time features of the control system. Vsystem is a network distributed control system software that provides distributed database access, and supports CAMAC I/O processing.

Two relational database products play an important role in the operation of the control system. The first relational database is Oracle's "Oracle Rdb" [6]. This database is used to store static information about accelerator parameters and devices that would be inappropriate for storage in the Vsystem real-time database. The second relational database is Corel's "Paradox" [7]. The Oracle Rdb and Paradox relational database systems team up to provide the operator with a graphical user interface to archived tune data. These stored data provide the operator with a means to restore the entire accelerator to a configuration used during a previous experiment.

---
[*] Undergraduate Research Participants

*2.3 System Personnel*

The control system staff most recently has consisted of one full-time system manager/software engineer, one full-time engineer, and two part-time undergraduate student programmers. This staff is responsible for maintaining all computer systems associated with the control system, the control system LAN, and the CAMAC subsystem. In addition, this group is responsible for all in-house written software maintenance, as well as any requests for new features and necessary upgrades.

## 3 MOTIVATIONS FOR MODIFYING THE CAMAC SUBSYSTEM DESIGN

While the CAMAC subsystem used at ATLAS has been both reliable and versatile, it has long been recognized that the system has the following inherent disadvantages:
- The serial highway configuration transmits commands and receives responses at one single point of contact. This design establishes a potential constraint to overall performance.
- CAMAC I/O processing is performed by only one computer. As the system grows, future peak demand periods could exceed the processing capability of this single machine.
- Operation of the entire control system depends on the reliability of one computer system. If this system is down, the entire control system is rendered inoperative.

## 4 ONE POSSIBLE APPROACH

It seems obvious that one solution to ease the burden of one computer system performing all of the CAMAC I/O, and to increase the number of I/O ports accessing CAMAC crates, is to add computer systems with CAMAC interfaces. Borrowing from the concepts of more contemporary control system designs such as the popular control system package EPICS (Experimental Physics and Industrial Control System) [8], local I/O processing could be added to selected CAMAC crates. While probably not practical at ATLAS, in principle, a single computer system could be interfaced directly to each of the 18 on-line CAMAC crates, thus providing local I/O processing for each crate.

Two things make this approach more feasible for the ATLAS control system today than in the past. The first is the low cost of PC hardware and software, and the second is the port of Vsystem to PC operating systems like Windows NT/2000 and Linux [9].

## 5 A TEST CASE

In the Vsystem environment, database records (Vsystem documentation refers to them as channels) contain all the information necessary to access the I/O hardware. At ATLAS, these records are organized primarily by system or device type into several different databases. Currently in the ATLAS control system, all Vsystem databases reside on the AlphaServer that is connected to the CAMAC serial highway. One of these databases is the cryogenics system database.

Due to the superconducting design of many ATLAS components, the reliability of monitoring cryogenic system parameters is crucial. Since the cryogenic system offered a good opportunity for funneling CAMAC I/O to one CAMAC crate, it was chosen as an ideal candidate for a prototype system.

## 6 PROTOTYPE IMPLEMENTATION

The following steps were taken to implement the prototype system:
- A PC was acquired to allow for a proof of principle investigation. The PC has been assembled in a rack mount configuration.
- A WIeNeR PCI to CAMAC interface and software driver was acquired for this initial machine [10].
- The Linux operating system and Vsystem were installed on the PC.
- The cryogenics database was then copied to the new PC, and modified to use the new PCI to CAMAC interface connected to the local CAMAC crate.
- The cryogenics graphical user displays were copied to the new PC. These displays, as well as the displays on the main system, were modified to reference the cryogenic database on the PC. This allowed data retrieved from the locally connected CAMAC crate to be displayed on all control system nodes.
- Cryogenics system signal cables were then moved from the serial highway crate to input modules in the new local test crate.

The result is that cryogenic system data can now be retrieved from a CAMAC crate that is not part of the CAMAC serial highway. These data can be displayed on the test PC or on any node that is part of the main system. The complete control system is illustrated in Figure 1.

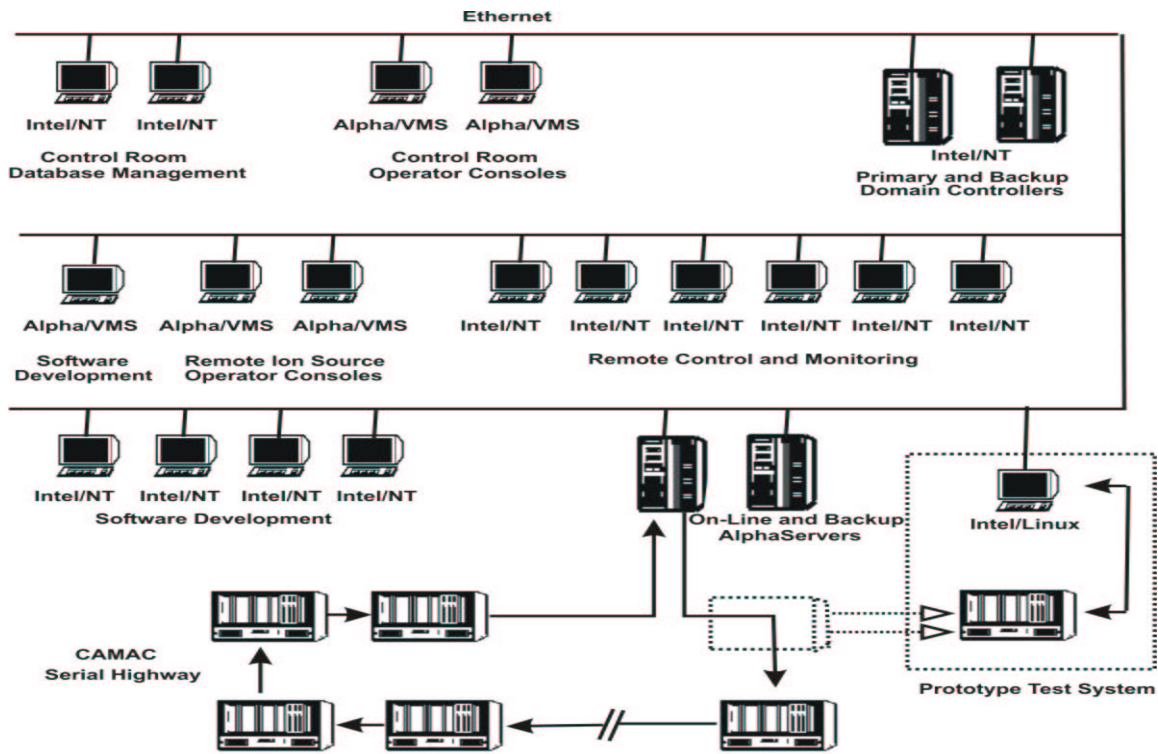

Figure 1: ATLAS Control System Showing Prototype Distributed I/O

## 7 CONCLUSIONS

It has been demonstrated, through initial experiences at ATLAS, that relatively low cost distributed CAMAC I/O processing is possible within the Vsystem environment. While converting the entire control system to a distributed configuration, as described in this paper, would be a major undertaking, it seems that the future of the ATLAS control system may no longer depend on the CAMAC serial highway as the only option. In fact, the prototype design suggests that the CAMAC subsystem itself is no longer the only option. The prototype system could just as easily have been configured to use a VME (Versa Module Europa) or possibly a VXI (VME eXtensions for Instrumentation) subsystem.

The current plan is to experiment with other bus structures, as well as other PC operating systems. This activity would allow ATLAS staff members to determine which would be most appropriate for the ATLAS control system.

This work is supported by the U.S. Dept. of Energy, Nuclear Physics Div., under contract W-31-109-ENG-38.

## REFERENCES

[1] Vista Control Systems, Inc., Los Alamos, NM, USA.
[2] F. Munson, D. Quock, B. Chapin, and J. Figueroa, "Argonne's ATLAS Control System Upgrade", International Conference on Accelerator and Large Experimental Physics Control Systems, ICALEPCS '99, Trieste, Italy, October 4-8, 1999.
[3] Compaq Computer Corporation, Houston, TX, USA.
[4] Kinetic Systems Corporation, Lockport, IL, USA.
[5] Microsoft Corporation, Redmond, WA, USA.
[6] Oracle Corporation, Redwood Shores, CA, USA.
[7] Corel Corporation, Jericho, NY, USA.
[8] Los Alamos National Laboratory, NM, and Argonne National Laboratory, IL, USA.
[9] Linux, A registered trademark of Linus Torvalds.
[10] WIeNeR, Plein & Baus, Burscheid, Germany.